\pdfoutput=1
\documentclass[a4paper]{article}
\usepackage{graphicx}
\usepackage{caption}
\usepackage{subcaption}
\usepackage{amsmath}
\usepackage[affil-it]{authblk}

\begin{document}
\title{Leaky-Wave Radiations by Modulating Surface Impedance on Subwavelength Corrugated Metal Structures}

\author[1]{Ben Geng Cai}
\author[1]{Yun Bo Li}
\author[1,2]{Hui Feng Ma}
\author[1,2]{Wei Xiang Jiang}
\author[1,3]{Qiang Cheng}
\author[1,3 *]{Tie Jun Cui}

\affil[1]{State Key Laboratory of Millimeter Waves, Department of Radio Engineering 
Southeast University, Nanjing 210096, China}
\affil[2]{Synergetic Innovation Center of Wireless Communication Technology, Southeast University, Nanjing, 210096, China}
\affil[3] {Cooperative Innovation Centre of Terahertz Science, No.4, Section 2, North Jianshe Road, Chengdu 610054, China}
\affil[*]{tjcui@seu.edu.cn}


\maketitle
%
%

\begin{abstract}
One-dimensional (1D) subwavelength corrugated metal structures has been described to support spoof surface plasmon polaritons (SPPs). Here we demonstrate that a modulated 1D subwavelength corrugated metal structure can convert spoof SPPs to propagating waves. The structure is fed at the center through a slit with a connected waveguide on the input side. The subwavelength corrugated metal structure on the output surface is regarded as metasurface and modulated periodically to realize the leaky-wave radiation at the broadside. The surface impedance of the corrugated metal structure is modulated by using cosine function and triangle-wave function, respectively, to reach the radiation effect. Full wave simulations and measuremental results are presented to validate the proposed design.
\end{abstract}


\section*{Introduction}
	
After the phenomenon of extraordinary transmissions through subwavelength holes was investigated by surface plasmon polaritons (SPPs) between metal and air in the optical region\cite{Ebbesen1998EOT, Lezec2002}, it was found that the corrugated metal structure on the output surface could shape the transmitted light emerging from the aperture\cite{Lezec2002, Martin2003PRL, Garcia2003APL}. The field on the exit surface has also been analyzed using the leaky surface-plasmon theory\cite{oliner_leaky_2003, jackson_leaky_2003, jackson_role_2008}. In the microwave region, a two-dimensional (2D) metal with drilling holes was proposed and investigated to support spoof SPPs\cite{pendry2004mimicking}. The one-dimensional (1D) corrugated metal structure has also shown to propagate spoof SPPs\cite{Garcia2005surfaces}. Similarly, electromagnetic beams shaped by the corrugated metal structure on the output surface\cite{beruete_enhanced_2004} have been observed, followed by the analysis of leaky-wave\cite{sutinjo2012simple} and the design of bull’s eye antennas in the microwave frequency\cite{baccarelli2004novel, lockyear2004surface, Beruete2005, liu2011coherent, beruete2013terahertz}, as has been done in the optical frequency\cite{Lezec2002, osorio2015k}. However, in the microwave region the period of the corrugated metal structures is about one wavelength to support leaky waves, and the angle of radiation beam was determined by the empirical formula\cite{beruete_enhanced_2004, Beruete2005} or complicated mode expansion method\cite{sutinjo2012simple}.

In recent years, subwavelength corrugations or grooves on metals supporting spoof SPPs with a period of approximately 1/6 wavelengths have been analyzed for slow-wave phenomena\cite{gan2008ultrawide, zhou2011broadband, shen2013conformal}. However, the radiations of leaky waves by these subwavelength corrugations have not been discussed. As spoof SPPs on the 1D structure is also a kind of transverse magnetic (TM) surface wave\cite{Garcia2005surfaces, sutinjo2012simple}, one could treat the subwavelength corrugations on metal as a kind of metasurface, the surface impedance of metasurface has been widely discussed and utilized to generate leaky waves, especially the sinusoidally modulated metasurfaces\cite{oliner1959guided, fong2010scalar, minatti2011spiral, patel2011printed, esquius2014sinusoidally, cai2015generation}. Conversely, the periodic structures satisfying the radiation condition could also be used to convert a propagating wave to surface wave\cite{sun2012gradient} due to the reversibility of light.

\begin{figure}[ht]
\centering
\includegraphics[width=12cm]{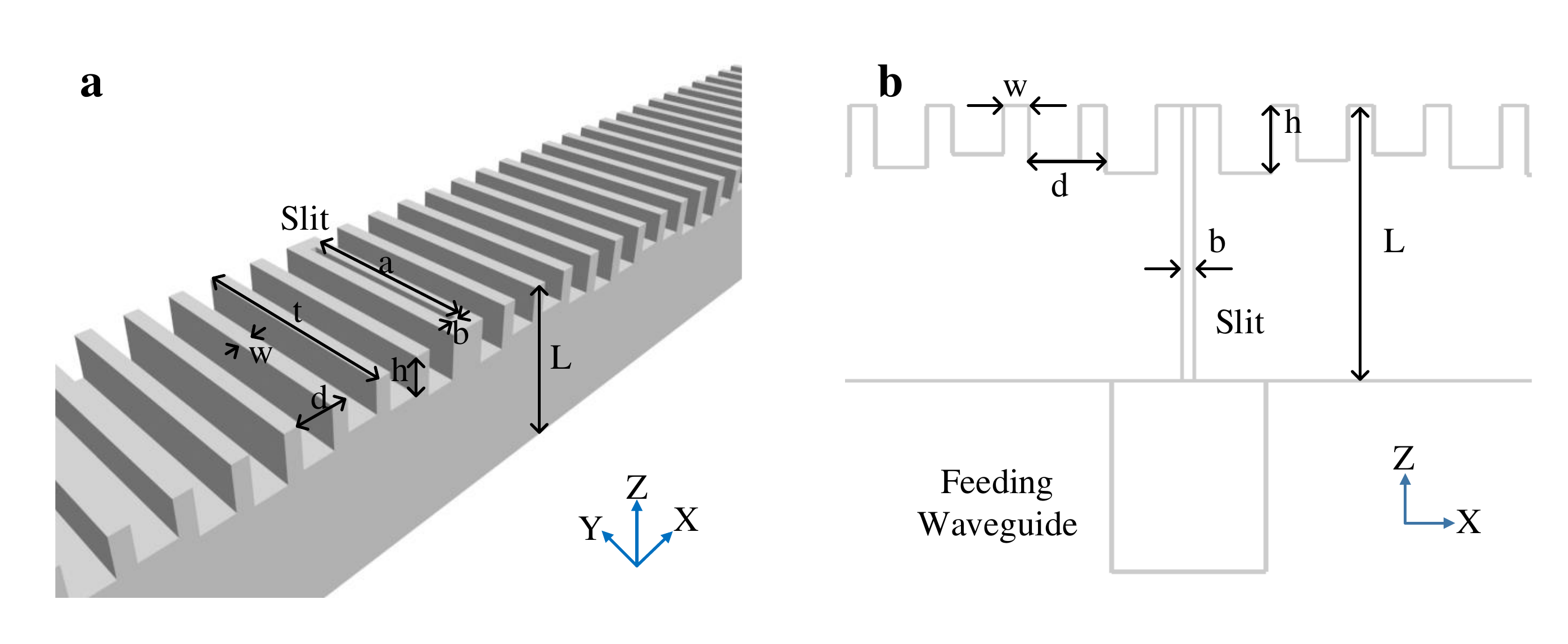}
\caption{ The schematic of subwavelength corrugated metal structure as a 1D periodic leaky-wave antenna. The center of the structure is a narrow slit which is connected to a feeding waveguide. The geometry parameters of the structure are given as: $w=1$ mm, $d=3$ mm, $h=2\sim 2.9$ mm, $t=15$ mm, $L=11.5$ mm, $a=12.5$ mm, $b=0.5$ mm, and $t=15$ mm. The waveguide has a size of 13mm$\times$6mm in the cross section and a longitudinal length of $8$ mm. The number of corrugations in each side from the slit is 70. (a) Perspective view of the corrugated metal structure. (b) Front view of the corrugated metal structure.}
\label{fig:Structure}
\end{figure}

Here we apply the concept of surface impedance to the subwavelength corrugated metal structure, which originally supports the spoof SPPs, and modulate the surface impedance using the cosine function and triangle-wave function, respectively, so that the modulated structures could radiate propagating waves as a periodic leaky-wave antenna. The antenna is fed at center of the structure through a slit, connecting to a feeding waveguide, as illustrated in Figure \ref{fig:Structure}. The energy is coupled from the slit to the subwavelength corrugated structure on the output surface of the metal to propagate spoof SPPs and radiate a narrow broadside beam with high directivity. The design process is analytical, and the experimental results agree well with the numerical simulations.

\section*{Results}

\subsection*{Theory and Design}

Like SPPs in the optical frequency, spoof SPPs is a kind of TM surface wave\cite{pendry2004mimicking, Garcia2005surfaces, gan2008ultrawide}. Subwavelength corrugated metal structures have been proposed and investigated to support spoof SPPs, where the dispersion relation has also been widely discussed. By treating the subwavlength corrugated metal structure as a metasurface, we could calculate the surface impedance of different unit cells easily by analyzing the dispersion diagrams of spoof SPPs. Then we could modulate the surface impedance of the corrugated structure periodically as a periodic leaky-wave antenna to radiate propagating waves from the spoof SPPs. Here we choose two typical functions, the cosine function and the triangle-wave function, to modulate the surface impedance in order to observe the broadside radiation of leaky waves.

The distribution of surface impedance corresponding to the subwavelength corrugated metal surface modulated with cosine function is written as
\begin{equation}
Z(x) = j[X + M \cos(nkx+kx\sin{\theta})],
\label{cosine}
\end{equation}
where $Z$ is the surface impedance along the $x$ direction that will be inductive for TM surface wave, $X$ is the average surface reactance, $M$ is the modulation depth, $k$ is the free-space wave number, $n=\sqrt{1+(X/Z_0)^2}$  represents the effective refractive index of surface wave, $Z_0=\sqrt{\mu_0/\epsilon_0}$ is the intrinsic impedance in free space, and $\theta$  is the polar angle of the radiation direction. The determination of the values for $X$ and $M$ will be discussed in the Methods. 

Here, we set $\theta = 0$  to observe a broadside radiation of leaky waves. The period of Eq. \eqref{cosine} is $p=2\pi/(nk)$, which is consistent with the theory of leaky waves for the first spatial harmonic:
\begin{equation}
\label{harmonic}
k_{-1}=nk-\frac{2\pi}{p}=0.
\end{equation}

For comparison, a triangle-wave function of surface impedance is also chosen, which is written in one period as:
\begin{equation}
\label{triangle}
    Z(x) = 
   \begin{cases}
   j(X + M -  \frac{4M}{p} x) &\mbox{if $0\leq x \leq p/2$}\\
   j(X + M +  \frac{4M}{p} x)  &\mbox{if $-p/2\leq x \leq 0$}
   \end{cases},
\end{equation} 
in which $X$, $M$ and period $p=2\pi/(nk)$ are the same as those in Eq. \eqref{cosine}.

A detailed structure of the antenna is illustrated in Fig. \ref{fig:Structure}, which is composed of a slit surrounded by metallic corrugations on the output side, and a feeding waveguide on the input side. The slit has a size of $a\times b$ in the cross section and a longitudinal length $L$. The energy is fed from a connected waveguide, which is tunneled into the slit. On the output side of the slit, the energy is coupled from the slit to the corrugated conductor which supports spoof SPPs or TM surface waves. Since the surface impedance of the corrugated conductor is modulated periodically with periodic functions mentioned above, a broadside radiation of leaky waves could be observed. 

The minimum reflection coefficient S11 at the designed frequency $f$ is determined by the Fabry-Perot resonance of TE$_{10}$ mode in the slit, leading to the optimal length of the slit, $L=11.5$mm, as shown in Fig. \ref{fig:Structure}.  The optimum process is described as follows. The dispersion relation of the TE$_{10}$ mode in the slit is given as $k^2=k_z^2 + k_y^2=k_z^2+{{\pi^2}/{a}}^2$, while the effective length  $L'$ of the Fabry-Perot resonance is a small modification of $L$ with  $L'=L+\triangle L$, which has to satisfy $k_z L'=\pi$ in order to realize the tunneling effect. Substituting $k_z L'=\pi$ into the dispersion relation, we have
\begin{equation}
\label{dispersion}
 k^2=k_z^2+k_y^2={\frac{\pi}{L'}}^2 + {\frac{\pi}{a}}^2,
\end{equation}
or
\begin{equation}
\label{frequency}
 f = \frac{c}{2}\sqrt{\frac{1}{L'^2} + \frac{1}{a^2}},
\end{equation}
where $c$ is the speed of light in vacuum. After that, for the designed frequency $f$, we could calculate the effective resonant length $L'$ as:

\begin{equation}
\label{length}
L' =\frac{1}{\sqrt{\frac{4f^2}{c^2} - \frac{1}{a^2}}},
\end{equation}
Hence, if the designed frequency is 17 GHz, the effective resonant length $L'$ is about $12.5$ mm, which is then set as the initial value for the optimum process. The length of slit is finally determined as $L=11.5$ mm, with a modification of $\triangle L=1$ mm.

\subsection*{Simulations and Experiments}

\begin{figure}[ht]
\centering
\includegraphics[width=8cm]{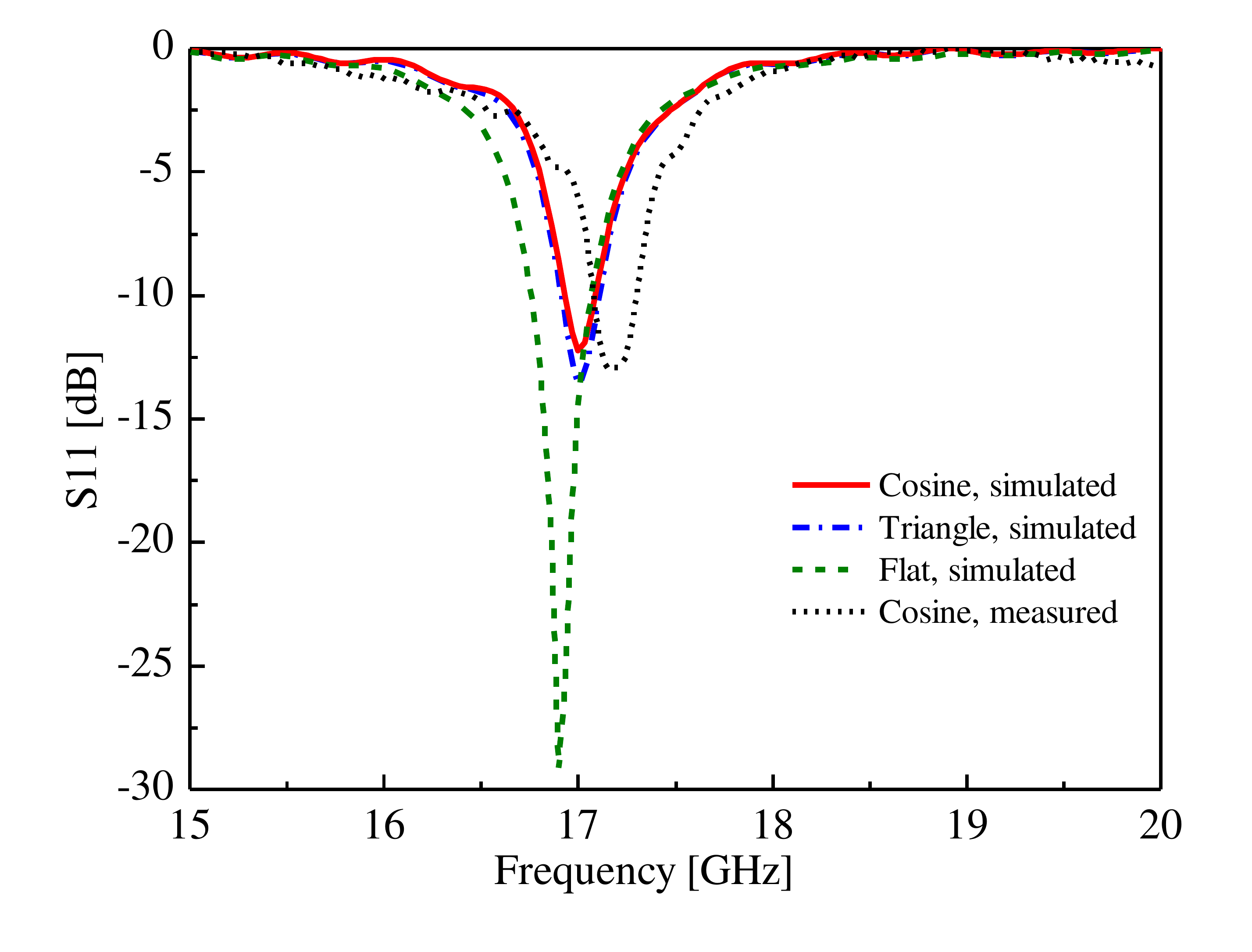}
\caption{The comparison of reflection coefficients (S11).}
\label{fig:s11}
\end{figure}

The simulated and measured reflection coefficients (S11) are shown in Fig. \ref{fig:s11}. The red solid curve corresponds to the simulation result of the cosine distribution for the corrugated structure, and the blue dashed-doted curve refers to the simulation result of triangle-wave function distribution for the corrugated structure. For comparison, the green dashed curve gives the simulation result of a flat metal structure without corrugation. The measured result of the corrugated structure with the cosine distribution is demonstrated as the black doted line. From Fig. \ref{fig:s11}, we clearly observe that both distributions in the cosine function and triangle-wave function have a minimum reflection coefficient (S11) under -10 dB near the designed frequency 17 GHz. For the flat structure, the minimum S11 has a small shift to the low frequency but the reflection coefficient at 17 GHz is still under -10 dB. The deviation is attributed to slightly different modification of $\triangle L$ as mentioned above. For the measured reflection coefficient of the cosine distribution, there is also a small frequency shift of 200 MHz, which might be caused from the differences between the fabrication, measurement and simulation.

\begin{figure}[ht]
\centering
\includegraphics[width=12cm]{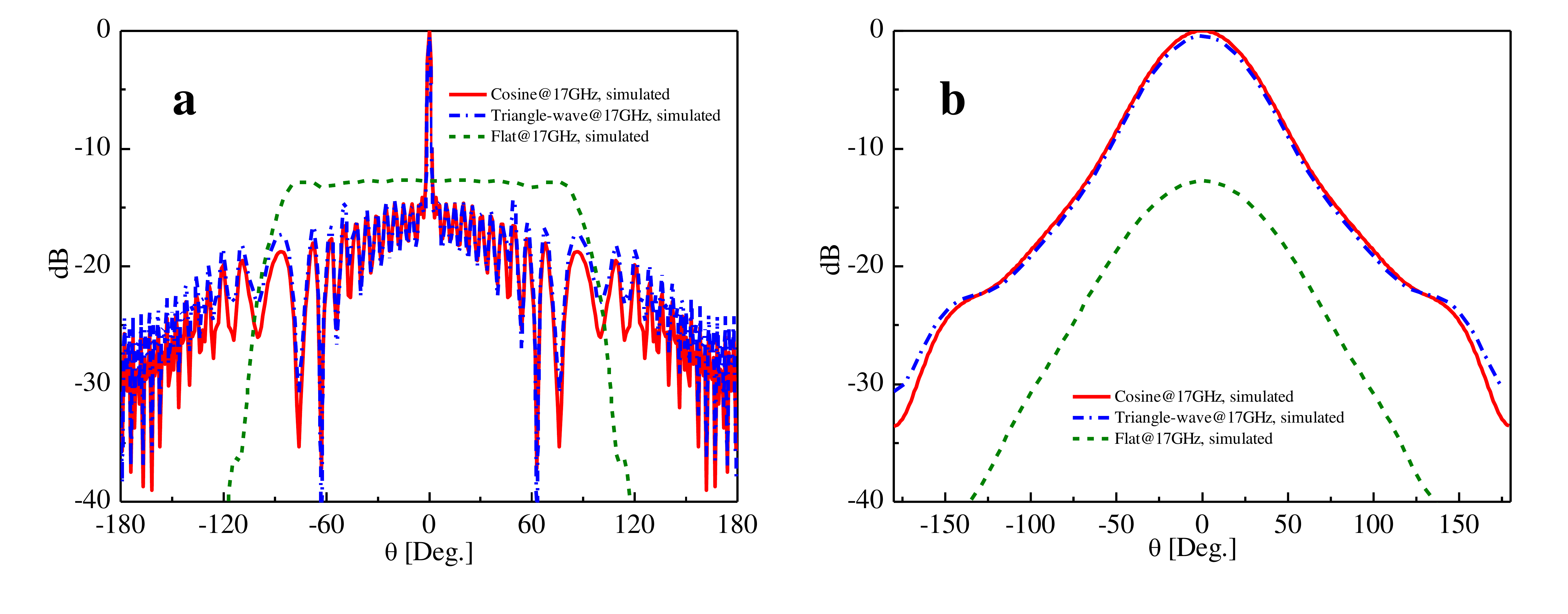}
\caption{The simulated normalized radiation patterns of the subwavelength corrugated metal structure and the flat conductor. (a) The E-plane radiation patterns. $\phi=0^\circ$ . (b) The H-plane radiation patterns. $\phi=90^\circ$. }
\label{fig:Simulated}
\end{figure}

\begin{figure}[ht]
\centering
\includegraphics[width=12cm]{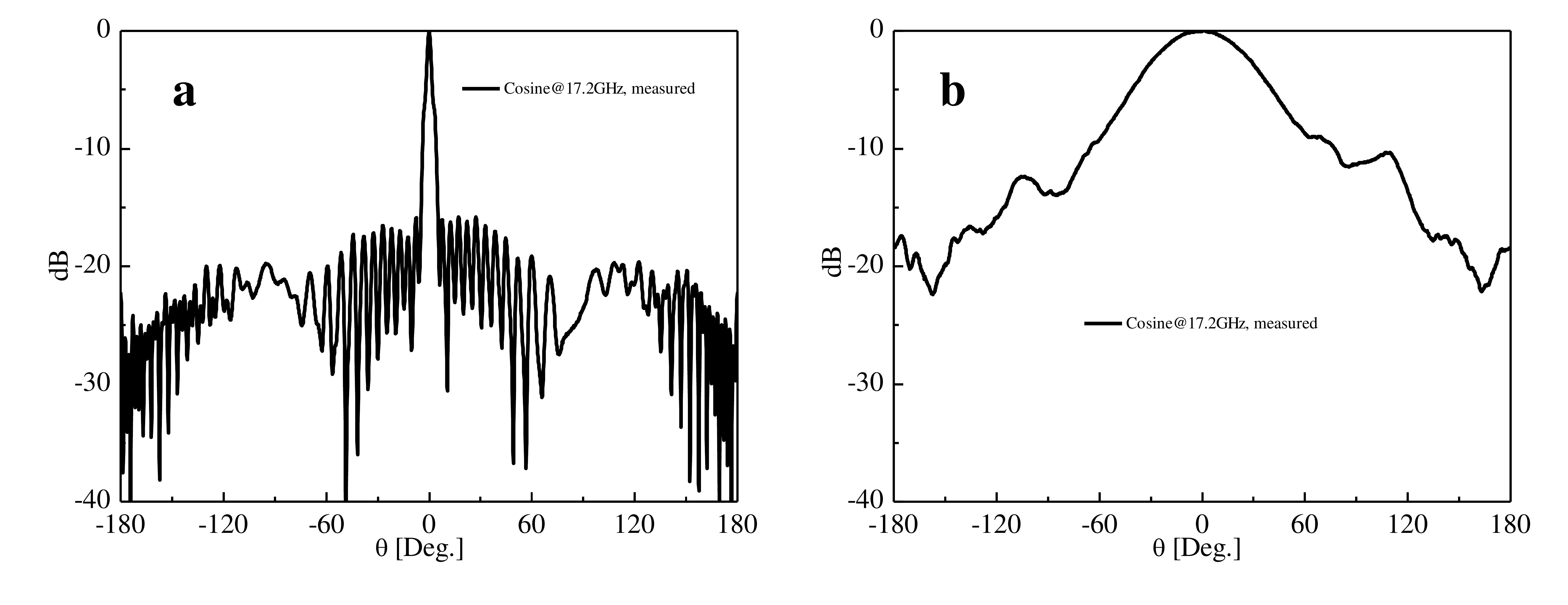}
\caption{The measured normalized radiation patterns of the subwavelength corrugated metal structure. (a) The E-plane radiation pattern. $\phi=0^\circ$ . (b) The H-plane radiation pattern. $\phi=90^\circ$. }
\label{fig:Measured}
\end{figure}

The normalized E-plane radiation patterns of the subwavelength corrugated metal structure and flat conductor at the designed frequency 17 GHz by simulations are illustrated in Fig. \ref{fig:Simulated}(a), from which we observe a broadside radiation in the cases of corrugated structures. Similar to Fig. \ref{fig:s11}, the red solid curve corresponds to the simulation result of the cosine distribution for corrugated structure, and the blue dashed-doted curve refers to the result of the triangle-wave function distribution. In both cases, the narrow -3 dB beam width is about $2.2^{\circ}$ and the side lobe levels are lower than -15 dB. For comparison, the green dashed curve gives the simulation result of the flat metal structure without corrugation. It could be observed that the maximum radiation enhancement for the corrugated structures in the E-plane is about 12.52 dB compared to the flat metal structure. The simulated gain of the corrugated structure at 17 GHz is 17.6 dB, while the simulated gain of the flat structure is about 5.08 dB. The normalized radiation patterns in the H-plane are also presented, as shown in Fig. \ref{fig:Simulated}(b).

The measured radiation patterns at 17.2 GHz are illustrated in Fig. \ref{fig:Measured}. Fig. \ref{fig:Measured}(a) shows the normalized E-plane radiation pattern of the cosine distribution for corrugated structure, from which we observe a narrow -3 dB beam width of $3^{\circ}$, and a side lobe level lower than -15 dB. The measured gain at 17.2 GHz is 16.67 dB, which is also close to the numerical results of 17.6 dB. Fig. \ref{fig:Measured}(b) shows the normalized H-plane radiation pattern correspondingly.

\section*{Discussion}
Conventionally, subwavelength corrugated metal structures have been studied for supporting spoof SPPs, but here we studied the radiation of leaky waves  using this structure. By treating the subwavelength corrugated conductor as metasurface to support TM surface waves and calculating the surface impedance of different unit cells, we then modulated the surface impedance periodically as periodic leaky wave antennas to radiate leaky waves to the desired directions. A broadside radiation is achieved near the designed frequency (17 GHz). It should be mentioned that this design is not limited to the broadside radiation, and the width $t$ could be adjusted freely since it is insensitive to the dispersion relations\cite{shen2013conformal}. The proposed  method could also be applied to metal disk to design a high gain bull's eye antenna\cite{beaskoetxea201577} or observe phenomena of pseudo Bessel beams\cite{cai2015generation} both in gigahertz and terahertz frequencies.

\section*{Methods} \label{Methods}

\begin{figure}[ht]
\centering
\includegraphics[width=12cm]{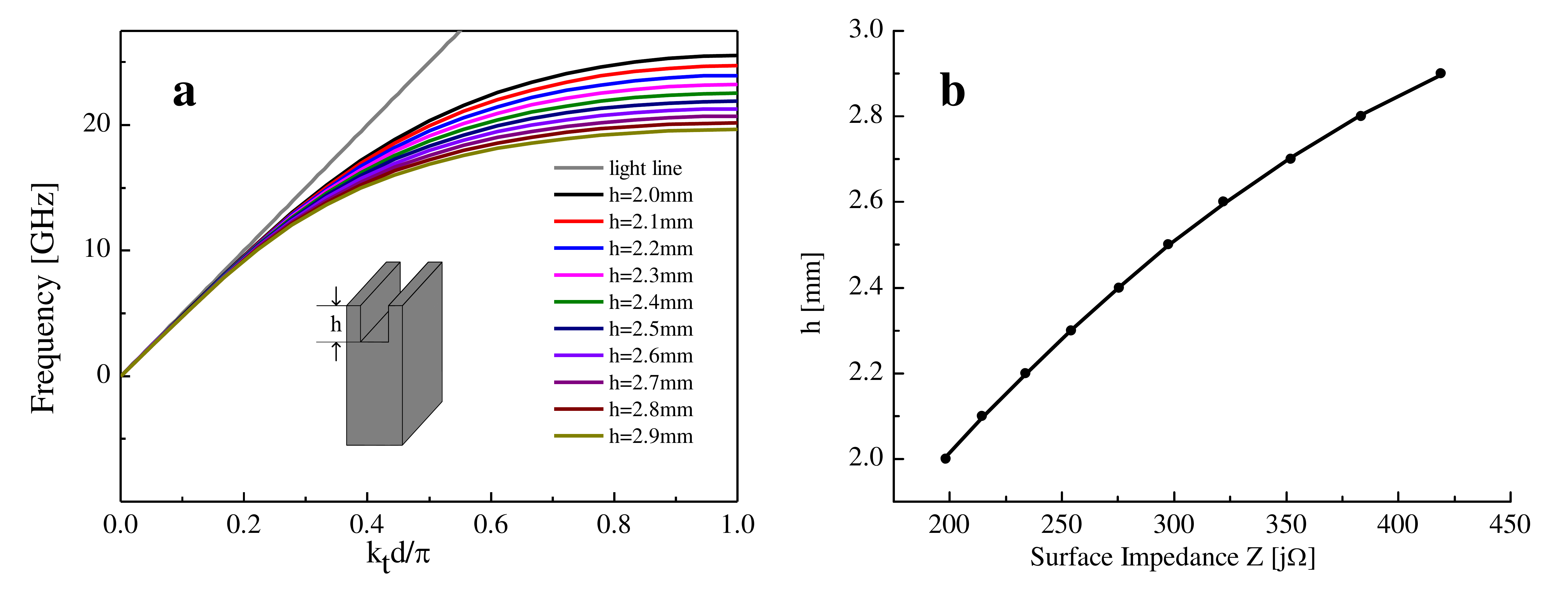}
\caption{The dispersion relations of the unit cells for different $h$ and the fitted curve for the surface impedance. (a) Dispersion relations of the unit cells. (b)  Surface impedances of different depths $h$.}
\label{fig:dispersion}
\end{figure}

Spoof SPPs propagating on the subwavelength corrugated metal structures are TM surface wave,  so that we can calculate the surface impedances of the corrugated conductors with different depths from the dispersion diagrams.  By simulating the unit cells of corrugated conductor with depths $h=2\sim2.9$ mm using the Eigenmode Solver in commercial software, CST Microwave Studio, we could get the dispersion diagrams of different unit cells, as shown in Fig. \ref{fig:dispersion}(a). Considering that the surface impedance is $Z=Z_0\sqrt{1-(k_t/k)^2}$, a function of $h=-7.634\times10^{-6}\times|Z|^2 + 0.008753\times|Z| + 0.569444$ is found by using the curve fitting, which can be used to design the distribution of the subwavelength corrugated metal structure, as illustrated in Fig. \ref{fig:dispersion}(b). The parameters in Eq. \eqref{cosine} are then calculated as $X=308.552$, $M=110.38$, $n=1.292$, and $\theta=0$.


\section*{Acknowledgements}

This work was supported in part from the National Science Foundation of China under Grant Nos. 61171024, 61171026, 61302018, 61401089,61571117, 61501112,  61522106, 61501117 and 61138001, in part from the 111 Project under Grant No. 111-2-05, in part from the Foundation of National Excellent Doctoral  Dissertation of China, and in part from the National Instrumentation Program under Grant No. 2013YQ200647.

\section*{Author contributions statement}
B.G.C and T.J.C. conceived the idea, analyzed the theoretical feasibility, and wrote the manuscript. B.G.C designed the samples and did the measurements. Y.B.L. and H.F.M. involved in the simulations and measurements. W.X.J. and Q.C. contributed to the discussions on theoretical feasibility and design improvements.

\section*{Additional information}

\textbf{Competing financial interests:} The authors declare no competing financial interests.

\end{document}